\documentstyle[12pt]{article}

\title{TRAPPED MODES IN THE VACUUM CHAMBER \\
 OF AN ARBITRARY CROSS SECTION}
\author{Sergey~S.~Kurennoy \\
        Physics Department, University of Maryland,
        College Park, MD 20742, USA \\
    and Gennady~V.~Stupakov \\ SLAC, Stanford, CA 94309, USA }

\begin{document}

\maketitle
\thispagestyle{empty}\pagestyle{empty}

\begin{abstract}
A recent study \cite{S&K} has shown that a small discontinuity
such as an enlargement or a hole on circular waveguides
can produce trapped electromagnetic modes with frequencies slightly
below the waveguide cutoff.
The trapped modes due to multiple discontinuities can lead to high
narrow-band contributions to the beam-chamber coupling impedance,
especially when the wall conductivity is high enough. To make
more reliable estimates of these contributions for real machines,
an analytical theory of the trapped modes is developed in this paper
for a general case of the vacuum chamber with an arbitrary
single-connected cross section. The resonant frequencies and
coupling impedances due to trapped modes are calculated, and simple
explicit expressions are given for circular and  rectangular
cross sections. The estimates for the LHC are presented.
\end{abstract}

\section{Introduction}

Previous computer studies of cavities coupled to a beam pipe indicated
that the impedance of small chamber enlargements exhibits sharp narrow
peaks at frequencies close to the cutoff frequencies of the waveguide,
see references in \cite{S&K}. For a single small discontinuity, such
as an enlargement or a hole, on a smooth circular waveguide, an
analytical theory has been developed \cite{S&K}, which shows that
these peaks can be attributed to trapped modes localized near the
discontinuity. A trapped mode is an eigenmode of the waveguide with a
discontinuity, with the eigenfrequency slightly below the waveguide
cutoff, which can exist in addition to the continuous
spectrum of the smooth waveguide.
The existence of trapped modes depends on a relation between
the conductivity of the chamber walls and a typical size of the
discontinuity, and in the limit of perfectly conducting walls the
trapped modes exist even for very small perturbations.

The trapped modes in a circular waveguide with many discontinuities
have also been studied \cite{SK94}, and it was demonstrated that
the resonance impedance due to $N$ close discontinuities in the
extreme case can be as large as $N^3$ times that for a single
discontinuity.
This phenomenon is dangerous for the beam stability in large
superconducting proton colliders like the LHC, where the design
anticipates a thermal screen (liner) with many small pumping holes
inside the beam pipe.
In such structures with many small discontinuities and a high
wall conductivity, the trapped modes can exist and contribute
significantly to the beam-chamber coupling impedances.

In the present paper we develop an analytical description of the
trapped modes for a waveguide with an arbitrary single-connected
cross section. We also derive particular results for circular
and rectangular waveguides from our general formulas.

\section{General Analysis}

Let us consider a cylindrical waveguide with a transverse cross
section $S$, having a small hole in its perfectly conducting walls.
We assume that the $z$ axis is directed along the waveguide axis,
the hole is located at the point ($\vec{b},z=0$), and its typical
size $h$ satisfies $h\ll b$. The fields of a source with time
dependence $\exp (-i \omega t)$ in the waveguide without hole
can be expressed as a series in TM- and TE-modes.
The fields of the TM${}_{nm}$ mode are \cite{Collin}
\begin{eqnarray}
E^\mp_z & = & k_{nm}^2 e_{nm} \exp(\pm \Gamma _{nm}z) \ ;
     \qquad  H^\mp_z = 0 \ ; \nonumber \\
\vec{E}^\mp_t & = & \pm \Gamma _{nm} \vec{\nabla}e_{nm}
        \exp(\pm \Gamma _{nm}z) \ ;               \label{emode} \\
Z_0 \vec{H}^\mp_t & = & ik \hat{z} \times \vec{\nabla}
   e_{nm} \exp(\pm \Gamma _{nm}z) \ , \nonumber
\end{eqnarray}
where $\pm$ indicates the direction of the mode propagation,
$k^2_{nm}$, $e_{nm}(\vec{r})$ are eigenvalues and orthonormalized
eigenfunctions (EFs) of the 2D boundary problem in $S$:
\begin{equation}
\left (\nabla ^2+ k^2_{nm}\right ) e_{nm} =
   0 \ ; \; \; e_{nm}\big\vert_{\partial S} = 0 \ ,  \label{boundpr}
\end{equation}
and propagation factors $\Gamma _{nm}=(k_{nm}^2-k^2)^{1/2}$ are to
be replaced by $-i \beta _{nm}$ with $\beta _{nm}=(k^2-k_{nm}^2)^{1/2}$
for $k>k_{nm}$.
Here $\vec{\nabla}$ is the 2D gradient in plane $S$; $k=\omega/c$;
$\hat{\nu}$ means an outward normal unit vector, $\hat{\tau}$ is
a unit vector tangent to the boundary $\partial S$ of the chamber
cross section $S$, and $\{ \hat{\nu},\hat{\tau},\hat{z}\}$ form a
RHS basis. Similarly, TE${}_{nm}$ fields are expressed in terms of
EFs $h_{nm}$ satisfying the boundary problem (\ref{boundpr}) with
the Neumann boundary condition $\nabla_\nu h_{nm}\vert_{\partial S}
= 0$, and corresponding eigenvalues $k'^2_{nm}$ \cite{Collin}.

\subsection{Frequency Shifts}

In the presence of the hole, there is a solution of the
homogeneous, i.e., without external currents, Maxwell equations
for this structure with the frequency $\Omega_s $
($s \equiv \{nm\}$) slightly below the corresponding cutoff
frequency $\omega_s= k_s c$, so that  $\Delta \omega_s \equiv \omega_s
 - \Omega_s \ll \omega_s$ --- the $s$th trapped TM-mode. At distances
$|z| > b$ from the discontinuity the fields of the trapped mode
have the form
\begin{eqnarray}
{\cal{E}}_z & = & k_s^2 e_s \exp(-\Gamma _s|z|) \ ;
     \qquad  {\cal{H}}_z = 0 \ ; \nonumber \\
\vec{\cal{E}}_t & = & \mbox{sgn}(z) \Gamma _s \vec{\nabla}e_s
        \exp(- \Gamma _s|z|) \ ;               \label{temode} \\
Z_0 \vec{\cal{H}}_t & = & ik \hat{z} \times \vec{\nabla}
   e_s \exp(- \Gamma _s|z|) \ , \nonumber
\end{eqnarray}
where $k = \Omega_s/c $, and the propagation constant $\Gamma _s$
satisfies the equation
\begin{equation}
\Gamma_s \simeq \frac {1}{4} \psi_{\tau} \left
 (\nabla_\nu e^h_s \right)^2   \ .  \label{GamE}
\end{equation}
Here $\psi_{\tau}$ is the transverse magnetic susceptibility of the
hole, cf.\ \cite{KGS95}, and superscript '$h$' indicates that the
field is taken at the hole. Typically, $\psi_{\tau}= O(h^3)$, while
$\nabla_\nu e^h_s = O(1/b)$, and as a result, $\Gamma_sb \ll 1$.
This means that the field of the trapped mode extends in the
waveguide over the distance $1/\Gamma_s$ large compared
to the waveguide transverse dimension.
Conditions like Eq.~(\ref{GamE}) were obtained in \cite{S&K,SK94}
for a circular waveguide using the Lorentz reciprocity theorem,
but there are other ways to derive them. For example, they follow
in a natural way from a general theory for the impedances of
small discontinuities \cite{KGS95}. In such a derivation, the
physical mechanism of this phenomenon becomes clear: a tangential
magnetic field induces a magnetic moment on the hole, and the induced
magnetic moment support this field if the resonance condition
(\ref{GamE}) is satisfied. Thus, the mode can exist even without an
external source, see in \cite{KGS95}. Note that the induced electric
moment $P_\nu$ is negligible for the TM-mode, since $P_\nu =
O(\Gamma_s)M_\tau$, as follows from Eq.~(\ref{temode}).

The equation (\ref{GamE}) gives the frequency shift $\Delta \omega_s$
of the trapped $s$th TM-mode down from its cutoff $\omega_s$
\begin{equation}
 \frac{\Delta \omega_s}{\omega_s} \simeq \frac{1}{32 k^2_s}
  \psi^2_{\tau} \left (\nabla_\nu e^h_s \right)^4 \ .  \label{dwE}
\end{equation}
In the case of a small hole this frequency shift is very small, and
for the trapped mode (\ref{temode}) to exist, the width of the
resonance should be smaller than $\Delta \omega_s $. Contributions to
the width come from energy dissipation in the waveguide wall due to its
finite conductivity, and from energy radiation inside the waveguide
and outside, through the hole. Radiation escaping through the hole is
easy to estimate \cite{S&K}, and for a thick wall it is exponentially
small, e.g., \cite{RLG}. The damping rate due to a finite conductivity
is $\gamma = P/(2W)$, where $P$ is the time-averaged power dissipation
and $W$ is the total field energy in the trapped mode, which yields
\begin{equation}
 \frac{\gamma_s}{\omega_s} = \frac{\delta}{4 k^2_s}
 \oint dl \left (\nabla_\nu e_s \right )^2 \ ,  \label{gamE}
\end{equation}
where $\delta$ is the skin-depth at frequency $\Omega_s$,
and the integration is along the boundary ${\partial S}$.
The evaluation of the radiation into the lower waveguide modes
propagating in the chamber at given frequency $\Omega_s$ is also
straightforward \cite{GS}, if one makes use of the coefficients of
mode excitation by effective dipoles on the hole, e.g., Eqs.~(6)-(9)
in Ref.\ \cite{KGS95}. It shows that corresponding damping rate
$\gamma_R = O(\psi^3_{\tau})$ is small compared to $\Delta \omega_s $.
For instance, if there is only one TE${}_p$-mode with the frequency
below that for the lowest TM${}_s$-mode, like in a circular waveguide
(H${}_{11}$ has a lower cutoff than E${}_{01}$),
\begin{equation}
 \frac{\gamma_R}{\Delta \omega_s} = \frac{\psi_\tau \beta'_p}{k'^2_p}
  \left (\nabla_\nu h^h_s \right )^2 \ ,  \label{gamER}
\end{equation}
where $\beta'_p \simeq (k_s^2-k'^2_p)^{1/2}$ because $k \simeq k_s$.

The frequency of the trapped TE${}_p$-mode is given
by the condition \cite{KGS95}
\begin{equation}
\Gamma'_p \simeq \frac{1}{4} \left [ \psi_{z} k'^2_p
 \left ( h^h_p \right)^2 - \chi \left
 (\nabla_\tau h^h_p \right)^2 \right ] \ ,      \label{GamH}
\end{equation}
provided the RHS of Eq.~(\ref{GamH}) is positive.
Here $\psi_z$ and $\chi$ are the longitudinal magnetic
susceptibility and the electric polarizability of the hole.

\subsection{Impedance}

The trapped mode (\ref{temode}) gives a resonance contribution to
the longitudinal coupling impedance at $\omega \approx \Omega_s$
\begin{equation}
 Z_s(\omega ) = \frac{2i \Omega_s \gamma_s R_s}
                     {\omega^2 - (\Omega_s -i\gamma_s)^2} \ ,
                   \label{imp}
\end{equation}
where the shunt impedance $R_s$ can be calculated as
\begin{equation}
R_s = {\sigma \delta  \left| \int{dz \exp (-i \Omega_s z /c)
{\cal E}_{z}(z)} \right|^2 \over \int_{S_w} ds
|{\cal H}_{\tau }|^2 } \ .
                 \label{Rimp}
\end{equation}
The integral in the denominator is taken over the inner wall surface,
and we assume here that the power losses due to its finite
conductivity dominate. Integrating in the numerator one should
include all TM-modes generated by the effective magnetic moment
on the hole using Eqs.~(6)-(9) from \cite{KGS95}, in spite of a
large amplitude of only the trapped TM${}_s$ mode. While all other
amplitudes are suppressed by factor $\Gamma_sb \ll 1$, their
contributions are comparable to that from TM${}_s$, because
this integration produces the factor $\Gamma_q b$ for any
TM${}_q$ mode. The integral in the denominator is dominated by
TM${}_s$. Performing calculations yields
\begin{equation}
R_s = \frac{Z_0 {\tilde{e}}^2_\nu \psi^3_{\tau} k_s
 \left (\nabla_\nu e^h_s \right)^4} {8 \delta
 \oint dl \left (\nabla_\nu e_s \right )^2  } \ ,   \label{Rs}
\end{equation}
where $\tilde{e}_\nu \equiv  - \sum_s e_s(0)
\nabla _{\nu}e_s(\vec{b})/k^2_s$ is the normalized electric field
produced at the hole location by a filament charge on the chamber
axis, see \cite{SK92} and \cite{KGS95}.

Results for a particular shape of the chamber cross section
are obtained from the equations above by substituting the
corresponding eigenfunctions.

\section{Circular Chamber}

For a circular cross section of radius $b$ the eigenvalues
$k_{nm}=\mu_{nm}/b$, where $\mu_{nm}$ is $m$th root of the
Bessel function $J_n(x)$, and the normalized EFs are
\begin{equation}
e_{nm}(r,\varphi) = \frac{J_n(k_{nm}r)}{\sqrt{N^E_{nm}}}
    \left \{ \begin{array}{c}
      \cos {n\varphi} \\ \sin {n\varphi} \end{array}
    \right \}   \ ,                          \label{ecnm}
\end{equation}
with $N^E_{nm} = \pi b^2 \epsilon_n J^2_{n+1}(\mu_{nm})/2$,
where $\epsilon_0=2$ and $\epsilon_n=1$ for $n \ne 0$.
For TE-modes, $k'_{nm}=\mu'_{nm}/b$ with $J'_n(\mu_{nm})=0$,
and
\begin{equation}
h_{nm}(r,\varphi) = \frac{J_n(k'_{nm}r)}{\sqrt{N^H_{nm}}}
    \left \{ \begin{array}{c}
      \cos {n\varphi} \\ \sin {n\varphi} \end{array}
    \right \}   \ ,                          \label{hcnm}
\end{equation}
where $N^H_{nm} = \pi b^2 \epsilon_n (1-n^2/\mu'^2_{nm})
J^2_n(\mu'_{nm})/2$. In this case $\tilde{e}_\nu = 1/(2\pi b)$,
which follows from the Gauss law. Assuming the hole located
at $\varphi = 0$, we get from Eq.~(\ref{GamE})
\begin{equation}
\Gamma_{nm} = \frac{\psi_\tau \mu^2_{nm}}
 {2 \pi \epsilon_n b^4 }    \ ,            \label{Gcnm}
\end{equation}
and from Eq.~(\ref{Rs})
\begin{equation}
 R_{nm} = \frac{Z_0 \psi^3_\tau \mu^3_{nm}}
 {32 \pi^4 \epsilon_n \delta b^8 }    \ .   \label{Rcnm}
\end{equation}
For TE-modes from Eq.~(\ref{GamH})
\begin{equation}
\Gamma'_{nm} = \frac{\psi_z \mu'^4_{nm}} {2 \pi
 \epsilon_n b^4 (\mu'^2_{nm}-n^2)} \ .    \label{G'cnm}
\end{equation}
Note that only the modes with $\cos {n\varphi}$ can be trapped,
while $\sin$-modes just do not ``see'' the hole.

The results of this section coincide with those of \cite{S&K,SK94},
except $R_s$ in \cite{S&K}, where the contribution of only the
trapped mode to Eq.~(\ref{Rimp}) was taken into account.
Formulas for an axisymmetric enlargement with area $A$ of the
longitudinal cross section are easily obtained from
Eqs.~(\ref{Gcnm})-(\ref{Rcnm}) with $n=0$ by substitution
$\psi_\tau \to 4\pi b A$.

\section{Rectangular Chamber}

For a rectangular chamber of width $a$ and height $b$ the
eigenvalues $k_{nm}=\pi \sqrt{n^2/a^2+m^2/b^2}$ for
$n,m = 1,2, \ldots$, and the normalized EFs are
\begin{equation}
e_{nm}(x,y) = \frac{2}{\sqrt{ab}}  \sin {\frac{\pi n x}{a}}
 \sin {\frac{\pi m y}{b}}  \ ,                  \label{ernm}
\end{equation}
with $0 \le x \le a$ and $0 \le y \le b$. Let a hole be located
in the side wall at $x=a, \ y=y_h$. Then Eq.~(\ref{GamE}) gives
\begin{equation}
\Gamma_{nm} = \frac{\psi_\tau \pi^2 n^2} {a^3 b}
 \sin^2 \left ( \frac{\pi m y_h}{b} \right ) \ ,    \label{Grnm}
\end{equation}
and from Eq.~(\ref{Rs}) the impedance is
\begin{equation}
 R_{nm} = \frac{Z_0 \psi^3_\tau \pi^3 n^2 \sqrt{n^2 b^2 + m^2 a^2}}
 {2 \delta a^4 b^2  (n^2 b^3 + m^2 a^3)}
 \Sigma^2 \left ( \frac{a}{b}, \frac{y_h}{b} \right )
 \sin^4 \left ( \frac{\pi m y_h}{b} \right )  \label{Rrnm}
\end{equation}
where
\begin{equation}
 \Sigma (u,v) = \sum_{l=0}^{\infty} \frac { (-1)^l \sin
 [\pi (2l+1) v ] }{ \cosh [\pi (2l+1) u / 2 ] }  \label{Sigma}
\end{equation}
is a fast converging series; see pictures in \cite{SK92}. Both
the frequency shift and especially the impedance decrease very
fast if the hole is displaced closer to the corners of the
chamber, i.e.\ when $y_h \to b$ or $y_h \to 0$.

\section{Estimates}

In a vacuum chamber with many discontinuities their mutual
interaction is very important. For trapped modes in a circular
pipe this interaction was studied in \cite{SK94}, but the results
are applicable for any cross section of the chamber. A few holes
in one cross section work as a single combined discontinuity.
If the average distance $g$ between adjacent cross sections with
holes is shorter than $1/\Gamma_s$, the number of the cross
sections with holes which work as an effective combined
discontinuity is $N_{eff} = \sqrt{2/(\Gamma_s g)}$. Referring to
\cite{SK94} for more detail, in this case we use the following
estimate for the reduced impedance of a cyclic accelerator
due to the trapped modes
\begin{equation}
Re \, \frac{Z}{n}  = \frac{4 \pi}
   {\Gamma_s k_s g^2} R_s \ ,          \label{ReZ/n}
\end{equation}
where $\Gamma_s$, $k_s$, and $R_s$ are given by the formulas
above.

For the LHC liner we consider a model having a square cross section
with side $a = 36$~mm. The liner wall has thickness $t=1$~mm and
the inner copper coating. There are 666 narrow longitudinal slots
with width $s=1.5$~mm and length $s=6$~mm per meter of the liner,
with $M=8$ slots in one cross section, which makes spacing
$g=12$~mm. The slots are located at distance $a/4$ from corners.
Using $\psi_\tau = w^2 s /\pi$ for a long slot in the thick wall
\cite{K&S}, we get for the lowest E-mode (TM${}_{11}$) near 5.9~GHz
\begin{equation}
Re \, \frac{Z}{n}  =  9.2 \sqrt{RRR} \mbox{ Ohms} \ , \label{LHC}
\end{equation}
where $RRR = 30 - 100$ for copper. The estimate for the model
with the circular cross section of radius $b = 18$~mm was
$16.5\sqrt{RRR}$~Ohms \cite{SK94}. These estimates presume identical
slots. A distribution of slot areas/lengths reduces $Re \, Z/n$
significantly: e.g., for the Gaussian distribution with RMS
$\sigma_A/A_{ave} = 0.1$, the above result $16.5\sqrt{RRR}$~Ohms
turns into 7~Ohms, independent of $RRR$, see \cite{SK94}.

\section{Conclusions}

The trapped modes in waveguides with an arbitrary single-connected
cross section are considered. The formulas for the frequency shift
and the resonance impedance are derived in a general case, and the
results for circular and rectangular chambers are given.

The transverse coupling impedance due to trapped modes is
calculated in a similar way, see formulas for the case of a circular
chamber in \cite{SK94}.

\end{document}